\documentclass[aps,prl,showpacs,twocolumn,amsmath,10pt,nofootinbib,superscriptaddress,floatfix]{revtex4-2}

\usepackage{graphicx,psfrag}
\usepackage{times}
\usepackage{bm} % for bold math symbols

\usepackage{amsmath,amsfonts,amssymb}
\usepackage{mathtools}
\usepackage{bbold}

\usepackage{multirow}
\usepackage{units}
\usepackage{enumerate}
\usepackage[normalem]{ulem} 
\usepackage{longtable}
\usepackage[rgb]{xcolor}
\usepackage{gensymb}
\usepackage{adjustbox}
\usepackage{enumitem}
\usepackage{hyperref}
\newcommand{\beq}{\begin{equation}}
\newcommand{\eeq}{\end{equation}}
\newcommand{\beqa}{\begin{eqnarray}}
\newcommand{\eeqa}{\end{eqnarray}}

\newcommand{\ben}{\begin{displaymath}}
\newcommand{\een}{\end{displaymath}}
\newcommand{\be}{\begin{equation}}
\newcommand{\ee}{\end{equation}}
\newcommand{\bea}{\begin{eqnarray}}
\newcommand{\eea}{\end{eqnarray}}
\newcommand{\nn}{\nonumber \\ }

\newcommand{\mprot}{m_{\rm p}}
\newcommand{\mneut}{m_{\rm n}}
\newcommand{\mn}{m_{\rm N}}
\newcommand{\fp}{f_{\rm p}}
\newcommand{\fn}{f_{\rm n}}
\newcommand{\fc}{f_{\rm c}}

\newcommand{\beginsupplement}{
  \setcounter{table}{0}
  \renewcommand{\thetable}{S\arabic{table}}%
  \setcounter{figure}{0}
  \renewcommand{\thefigure}{S\arabic{figure}}%
  \setcounter{equation}{0}
  \renewcommand{\theequation}{S\arabic{equation}}%
}

\begin{document}
%\begin{bibunit}[apsrev4-2]
%\nocite{apsrev42Control}

\title{Precision determination of pion-nucleon coupling constants 
  using effective field theory} 

\author{P.~Reinert}
%\email[]{patrick.reinert@rub.de}
\affiliation{Ruhr-Universit\"at Bochum, Fakult\"at f\"ur Physik und Astronomie,
Institut f\"ur Theoretische Physik II,
D-44780 Bochum, Germany}
\author{H.~Krebs}
%\email[]{hermann.krebs@rub.de}
\affiliation{Ruhr-Universit\"at Bochum, Fakult\"at f\"ur Physik und Astronomie,
Institut f\"ur Theoretische Physik II,
D-44780 Bochum, Germany}
\author{E.~Epelbaum}
%\email[]{evgeny.epelbaum@rub.de}
\affiliation{Ruhr-Universit\"at Bochum, Fakult\"at f\"ur Physik und Astronomie,
Institut f\"ur Theoretische Physik II,
  D-44780 Bochum, Germany}

\begin{abstract}
The pion-nucleon coupling constants determine the strength of the
long-range nuclear forces and play a fundamental part in our
understanding of nuclear physics.
While the charged- and neutral-pion
couplings to protons and neutrons are expected to be very similar,
owing to the approximate isospin symmetry of the strong interaction, the
different masses of the up- and down-quarks and electromagnetic
effects may result in their slightly different values.
Despite previous attempts to extract these
%pion-nucleon
coupling constants
% charged- and neutral-pion
%couplings to protons and neutrons
from different systems,
our knowledge of their values is still deficient.
In this Letter we present a precision determination of these
fundamental observables with fully controlled uncertainties 
from neutron-proton and proton-proton scattering data using chiral effective field
theory. To achieve this goal, we use a novel
methodology based on the Bayesian approach and perform, for the first
time, a full-fledged partial-wave analysis of nucleon-nucleon
scattering up to the pion production threshold in the framework of chiral effective field theory, including a complete
treatment of isospin-breaking effects and our own determination of mutually consistent data. The resulting values of the
pion-nucleon coupling constants are accurate at the percent
level and show no significant charge dependence. These results mark an
important step towards developing a precision theory of nuclear forces
and structure.
\end{abstract}


\maketitle
The pion-exchange mechanism drives the low-energy interaction between
protons and neutrons and is of utmost importance for our
understanding of atomic nuclei that make up $99.9\%$ of the visible
universe and for answering big science questions such as the origin of
the elements, the limits of nuclear stability, searches for physics
beyond the Standard Model and physics of neutron stars.
The interaction of
%pions and nucleons
a charged and neutral pion with protons (p) and
neutrons (n)
is characterized by three coupling constants $f_{\pi^0 \rm pp}$,
$f_{\pi^0 \rm nn}$ and $f_{\pi^\pm \rm pn}$, whose precise definitions
will be given below. These three constants 
determine the strength of the long- and intermediate-range nuclear
forces originating from exchange of virtual pions. Their precise knowledge with controlled uncertainties is,
therefore, of fundamental importance for a quantitative
understanding of nuclear physics. 

While possible
in principle, the \emph{ab initio} precision determination of the
pion-nucleon ($\pi$N) coupling constants 
from lattice  quantum-chromodynamics (QCD) and quan\-tum-electrodynamics (QED) calculations is presently out of
reach \cite{Bali:2019yiy}. Attempts have been made to extract $f_{\pi^\pm \rm pn}$ from experimental data on pion-nucleon ($\pi$N) scattering
\cite{Koch:1980ay,MarkopoulouKalamara:1993rv,Arndt:2003if}, 
pionic atoms \cite{Ericson:2000md,Baru:2011bw} and 
proton-antiproton scattering \cite{Timmermans:1994pg},
but the best way to determine all three
% coupling
constants is by
analyzing the abundance of proton-proton (pp) and neutron-proton (np) scattering
data. However, previous studies along this line
\cite{Klomp:1991vz,Stoks:1992ja,Bugg:1994ph,deSwart:1997ep,Perez:2016aol}
relied on
phenomenological models and offered no way of a reliable uncertainty
quantification apart from estimating statistical errors.

In this
Letter we
% , for the first time,
determine the values of all three $\pi$N
coupling constants from nucleon-nucleon (NN) scattering data using the
model-independent framework of chiral effective field theory
(EFT) \cite{Weinberg:1990rz,Epelbaum:2008ga}. This method has already been
% extensively
applied to NN
scattering, and the EFT expansion of the NN force has been recently 
pushed to fifth order 
(N$^4$LO) \cite{Epelbaum:2014sza,Entem:2017gor,Reinert:2017usi}.
% See \cite{} for selected review articles.
The crucial new aspects of
the current investigation include:

(i) For the first time, a determination of all three $\pi$N coupling
constants and a partial wave
analysis of NN data up to the pion production threshold including the
determination of mutually consistent data are carried out in the framework of chiral EFT.
We have taken into account
%extended the chiral EFT interactions of
%\cite{Reinert:2017usi} by taking into account
\emph{all} 
charge-independence-bre\-a\-king (CIB) and
charge-symmetry-breaking (CSB) isospin-vi\-o\-la\-ting NN interactions up through
N$^4$LO. This allowed us to achieve
% , for the first time,
a statistically perfect description of mutually consistent pp and
np scattering data
% up to the pion production threshold
in
the framework of chiral EFT that is unprecedented in its precision.

(ii) We have succeeded in overcoming the computational challenge of
performing a {\it Bayesian} determination of the $\pi$N coupling
constants. Contrary to the computationally much less demanding
frequentist methods used in all previous determinations
\cite{Klomp:1991vz,Stoks:1992ja,Bugg:1994ph,deSwart:1997ep,Perez:2016aol,Rentmeester:1999vw},
the Bayesian approach provides a {\it rigorous} way to calculate the joint conditional 
probability density of the  $\pi$N coupling constants given NN data.
%Our analysis does, in particular, not rely on the multivariate normal assumption for the 
%probability density function (PDF) of the involved parameters.

(iii) A careful uncertainty analysis, facilitated by the usage of
Bayesian methods, is performed to estimate not only
statistical errors in the calculated $\pi$N coupling
constants, but also systematic uncertainties from the truncation of
the EFT expansion and the choice of the highest energy of the included
NN data -- a feature lacking in the earlier determinations
of these quantities.

{\it Definitions of the $\pi$N coupling constants.}---Consider first 
the interaction of the nucleon with the
isovector weak current $A_i^\mu$, which 
is described in terms of the axial and induced pseudoscalar form
factors $G_A$ and $G_P$. In the limit of exact isospin
symmetry corresponding to the equal masses of the up- and down-quarks
and in the absense of electromagnetic interactions, the matrix element of $A_i^\mu (x = 0)$ between nucleon states
can be parametrized via  
\begin{displaymath}
\langle {\rm N} (p ') | A^\mu_i (0) | {\rm N} (p) \rangle = \bar u (p ') \left[
  \gamma^\mu G_A  + \frac{q^\mu}{2 \mn } G_P 
\right] \gamma_5 \frac{\tau_i}{2} u(p)\,,
\end{displaymath}
with $u(p)$ and $\bar u (p')$ the corresponding Dirac spinors, $\tau_i$
the isospin Pauli matrices and $\mn$ the nucleon mass. Furthermore, $q^\mu
= (p' - p)^\mu$ refers to the momentum transfer of the nucleon.
The form factors $G_A(q^2)$ and $G_P(q^2)$ carry important information about the internal
structure of the nucleon. For example, the axial
charge of the nucleon $g_A \equiv G_A (0) = 1.2756(13)$ \cite{Zyla:2020zbs} controls the decay rate of a
neutron to a proton.  Recently, this quantity was calculated from
first principles at a per-cent level using lattice
lattice quantum-chromodynamics (QCD) \cite{Chang:2018uxx}, see also \cite{Aoki:2019cca} for a review of lattice QCD
calculations of $g_A$.   While $G_A(q^2)$ is a smooth
function near $q^2 = 0$, the induced pseudoscalar form factor
possesses a pion-pole contribution, $G_P(q^2) = 4 \mn
g_{\pi \rm NN} F_\pi/(M_\pi^2 - q^2) + \mbox{non-pole
  terms}$, whose residue is determined by
the (pseudoscalar) $\pi$N coupling constant $g_{\pi \rm NN}$. The pion decay
constant $F_\pi= (92.1 \pm 0.8)$~MeV \cite{Zyla:2020zbs} determines the rate of weak decays $\pi^\pm \to \mu^\pm \nu_\mu$.
The strong-interaction constant $g_{\pi \rm NN}$ is connected to
$g_A$ and $F_\pi$ entering weak processes via the
celebrated Goldberger-Treiman relation $F_\pi g_{\pi \rm NN} = g_A \mn
(1 + \Delta_{\rm GT})$, where the small Goldberger-Treiman discrepancy
$\Delta_{\rm GT}$ is driven by the non-vanishing masses of the up- and
down-quarks.

Away from the isospin limit and in the presence of
quan\-tum-electrodynamics (QED), one has to distinguish
between protons  and neutrons and between the charged and neutral
pions by introducing three coupling constants $g_{\pi^0 \rm pp}$,
$g_{\pi^0 \rm nn}$ and $g_{\pi^\pm \rm pn}$ or, equivalently, the
corresponding pseudovector couplings $\fp \equiv f_{\pi^0 \rm pp} = M_{\pi^\pm}
g_{\pi^0 \rm pp} /(2 \sqrt{4 \pi} \mprot )$, $\fn \equiv f_{\pi^0 \rm nn} = M_{\pi^\pm}
g_{\pi^0 \rm nn} /(2 \sqrt{4 \pi} \mneut )$ and $\fc \equiv f_{\pi^\pm \rm pn} = M_{\pi^\pm}
g_{\pi^\pm \rm pn} /(\sqrt{4 \pi} (\mprot + \mneut ))$. The
determination of these fundamental constants from NN scattering data is the
main subject of this study.

\begin{figure}[tb]
  \begin{center}
\includegraphics[width=0.49\textwidth,keepaspectratio,angle=0,clip]{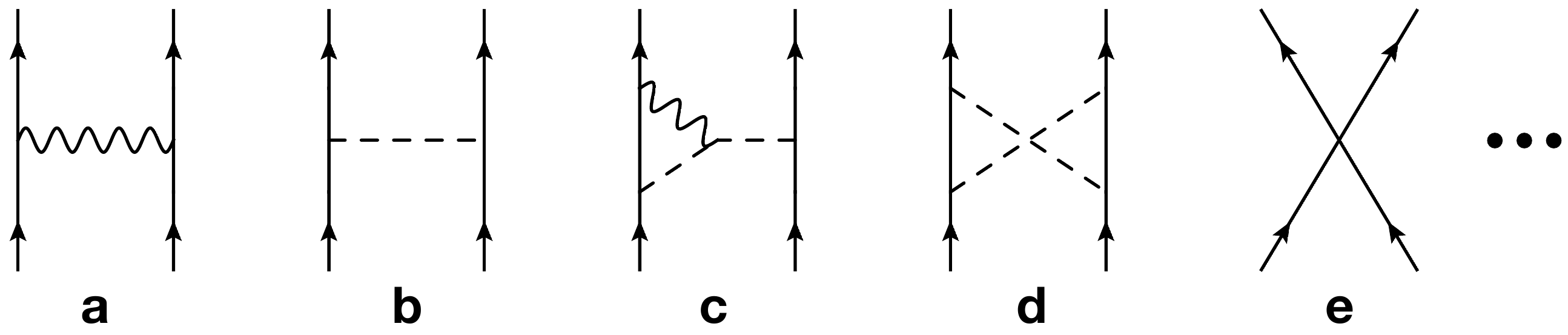}  
\end{center}
\vskip -0.4 true cm 
\caption{Diagrammatic illustration of the NN interaction in
    chiral EFT. Photons, pions and nucleons are shown by wavy,
  dashed and solid lines, respectively. Diagrams \textbf{a},
  \textbf{b}, \textbf{c}, \textbf{d} and \textbf{e} are representative
  examples of the one-photon exchange, one-pion exchange (OPE), electromagnetic corrections
  to the OPE, two-pion exchange (TPE) and NN short-range contributions,
  respectively. The range of interactions decreases from the left to
  the right.  \label{fig1}
}
\end{figure}

{\it Chiral EFT for nuclear forces.}---
We use chiral EFT, an effective field theory of QCD,  to describe the low-energy
interactions between two nucleons, and employ the resulting NN
potential to extract the $\pi$N coupling constants from a combined
Bayesian analysis of np and pp scattering data below the pion production
threshold. Chiral EFT utilizes an expansion in powers of momenta and pion masses 
to describe
% low-energy
interactions between pions and nucleons
in a systematically improvable
way \cite{Weinberg:1990rz,Bernard:1995dp,Epelbaum:2008ga}.
%, see refs.~\cite{Bernard:1995dp,Epelbaum:2008ga} for
%reviews.
The corresponding effective Lagrangian contains all possible terms
compatible with the symmetries of QCD. The non-perturbative dynamics
of QCD is encoded in the so-called low-energy constants (LECs), which
control the strength of the interactions in the effective Lagrangian and can be
determined from experiments or lattice QCD calculations. Chiral EFT
has also been extended to include virtual photons.
%in Ref.~\cite{Urech:1994hd}. 
Fig.~\ref{fig1} shows examples of contributions to the 
NN force in chiral EFT.  
The most important terms at leading order (LO)
include one-pion exchange (Fig.~\ref{fig1}b) and contact interactions
(Fig.~\ref{fig1}e). Two-pion exchange (Fig.~\ref{fig1}d) and
one-photon exchange (Fig.~\ref{fig1}a)
start to contribute at next-to-leading order (NLO), while pion-photon
exchange (Fig.~\ref{fig1}c) appears first at fourth order
(N$^3$LO). 

In recent years, the chiral expansion of the NN force has been pushed to
fifth order \cite{Epelbaum:2014sza,Entem:2017gor,Reinert:2017usi}.
%see refs.~\cite{Epelbaum:2019kcf,RodriguezEntem:2020jgp} for review articles.
All relevant isospin-invariant $\pi$N LECs have been reliably determined
from a dispersion theory analysis of $\pi$N scattering in
Ref.~\cite{Hoferichter:2015tha}. Therefore, the long-range part of
the NN interaction is parameter-free.
To avoid distortions of the 
long-range forces due to a finite cutoff $\Lambda$,
%which in the employed theoretical framework has to be kept finite,
we introduced in Ref.~\cite{Reinert:2017usi} an 
improved local regulator which
respects the analytic structure of the interaction.
The LECs accompanying short-range operators
(Fig.~\ref{fig1}e) were determined in Ref.~\cite{Reinert:2017usi} from
a fit to the 2013 Granada database \cite{Perez:2013jpa} of
mutually compatible np and pp data. Furthermore, we have introduced a
N$^4$LO$^+$ NN potential, where the leading  F-wave short-range
interactions, formally appearing at sixth order, were taken
into account in order to achieve a statistically satisfactory
description of certain very precisely measured pp data, see
also Ref.~\cite{Entem:2017gor}. This allowed us to achieve a
description of NN data on par with 
or even better than that based on the most precise phenomenological
potentials, but with
a much smaller number of adjustable parameters. However, the treatment
of isospin-breaking (IB) effects in Ref.~\cite{Reinert:2017usi} was
incomplete
% from the point of view of chiral EFT
and limited
to the one of the Nijmegen \cite{Stoks:1993tb} and Granada 2013 \cite{Perez:2013jpa}
partial wave analyses (PWA). 

In this Letter we
% , for the first time,
include the charge-independence-breaking (CIB) and
charge-symmetry-breaking (CSB) IB NN interactions complete up through
N$^4$LO. In particular, we employ the most general form of the
OPE potential including the leading electromagnetic
corrections \cite{vanKolck:1997fu} and take into account 
the leading and subleading IB two-pion-exchange
contributions \cite{Friar:1999zr,Friar:2003yv,Epelbaum:2005fd}.  These long-range interactions are 
expressed in terms of known LECs, the $\pi$N coupling
constants $\fp^2$, $\fc^2$ and  $f_0^2 \equiv \fp \fn$ to be determined,
the nucleon mass difference $\delta m = \mneut - \mprot
\simeq 1.29$~MeV  and its QCD contribution
$\delta m^{\rm QCD} = 2.05(30)$~MeV \cite{Gasser:1974wd},
see Ref.~\cite{Gasser:2020mzy} for an update and Ref.~\cite{Borsanyi:2014jba} for a 
recent {\it ab initio} calculation using lattice QCD and QED.
We also include short-range IB interactions in the $^1$S$_0$, $^3$P$_0$,
$^3$P$_1$ and $^3$P$_2$ partial waves.
Details of the employed NN interaction are given in Supplemental
Material.
% and the corresponding LECs need to be
% extracted from NN data.

{\it Determination of the $\pi$N coupling constants.}---We end up with $33$
parameters that need to be determined from NN data, comprising of 3 $\pi$N LECs $f^2 \equiv
\{ \fc^2, \, \fp^2,  \, f_0^2\}$ and $25 + 5$ LECs $C_i$ from
isospin-invariant $+$ IB short-range interactions,
collectively denoted as $C \equiv \{C_i\}$. For normally
distributed errors, the  likelihood of data $D$ given $f^2$, $C$ and
$\Lambda$ is given by
\begin{equation}
    \label{eq:likelihood}
    p(D \vert f^2 C \Lambda) = \frac{1}{N} e^{-\frac{1}{2}\chi^2},
  \end{equation}
  where $N$ is a normalization constant.
The data $D$ employed in our analysis include mutually compatible np and pp scattering data
according to our own selection
as detailed in Supplemental Material, where we also
provide the definition of
  the $\chi^2$-measure.
Using Bayes' theorem to relate the probability density function (PDF)
$p(f^2 C \Lambda \vert D)$  of the parameters given the data to $p(D \vert f^2 C
  \Lambda)$, and integrating over  the 
  nuisance parameters $C$ and $\Lambda$, we obtain the
quantity we are actually interested in, namely
the PDF of $f^2$
given the data $D$
\begin{equation}
    \label{eq:OPEPcouplingsPDF}
    p(f^2 \vert D) = \int d\Lambda \, dC \; \frac{p(D \vert f^2 C \Lambda) p(f^2 C \Lambda)}{p(D)}\,.
  \end{equation}
For the case at hand,
$p(D)$ is a (normalization) constant. 
  Furthermore, we use independent priors for $f^2$, $C$ and $\Lambda$ so
  that $p(f^2 C \Lambda) = p(f^2) \, p(C) \, p(\Lambda)$, and employ a
  Gaussian prior for $C$ and uniform priors  for $\Lambda$ and $f^2$
  specified in Supplemental Material.
  To determine $f^2$ we need to find the maximum of $p(f^2 
\vert D)$ in Eq.~\eqref{eq:OPEPcouplingsPDF}. However, for each set of $f^2$,
this requires integrating over a $31$-dimensional space spanned by $\Lambda$ and $C$,
which is not feasible. Instead, we employ the Laplace approximation 
by fitting  $C$ to $D$ for fixed values of $f^2$ and
$\Lambda$ and expressing 
the likelihood $p(D \vert f^2 C\Lambda)$ as 
\begin{equation}
    \label{eq:likelihood-approximation}
    p(D \vert f^2 C \Lambda) \approx \frac{1}{N} e^{-\frac{1}{2}[\chi^2_{\rm min} + \frac{1}{2}(C-C_{\rm min})^T H (C-C_{\rm min})]}\,. 
  \end{equation}
 Here,
$\chi^2_{\rm min} \equiv \chi^2_{\rm min}(f^2,\Lambda)$ at $C_{\rm
  min} \equiv C_{\rm min}(f^2,\Lambda)$ and the Hessian $H \equiv
H(f^2,\Lambda)$ is given by $H_{ij}= \frac{\partial^2 \chi^2}{\partial
  C_i \partial C_j}\big|_{C=C_{\rm min}}$. Performing an analytical integration over $C$
then allows us to cast 
Eq.~\eqref{eq:OPEPcouplingsPDF} into a numerically tractable form, see
Supplemental Material for details. The remaining integration over $\Lambda \in
[400,
\, 550]$~MeV is
performed numerically.
We emphasize that reducing the amount of information in the
employed priors for $C$, $f^2$ and $\Lambda$ has a 
negligible effect on our results, see Supplemental Material for details.

To account for the uncertainty inherent in the choice of the
energy range of our PWA, we performed separate analyses of NN data up to the laboratory
energies of $E_\mathrm{lab}^\mathrm{max} = 220$, $240$, $260$, $280$ and
$300$~MeV. Furthermore, to address the systematic error stemming from the truncation
of the EFT expansion for IB interactions, we considered two additional
models of the NN interaction
that include IB pion-photon- and two-pion-exchange contributions beyond
N$^4$LO. Our final PDF for the $\pi$N coupling constants are obtained 
by performing
% Bayesian
averaging over five values for $E_\mathrm{lab}^\mathrm{max}$ and three
models for IB interactions in order to account for the
  truncation
uncertainty at N$^4$LO as detailed in Supplemental Material.  
For all considered cases, the self-consistency of our results is verified by
comparing the quantiles of the $\chi^2$ residuals with those of the assumed
normal distribution (see Supplemental Material). 
\begin{figure}[tb]
  \begin{center}
    \includegraphics[width=0.49\textwidth,keepaspectratio,angle=0,clip]{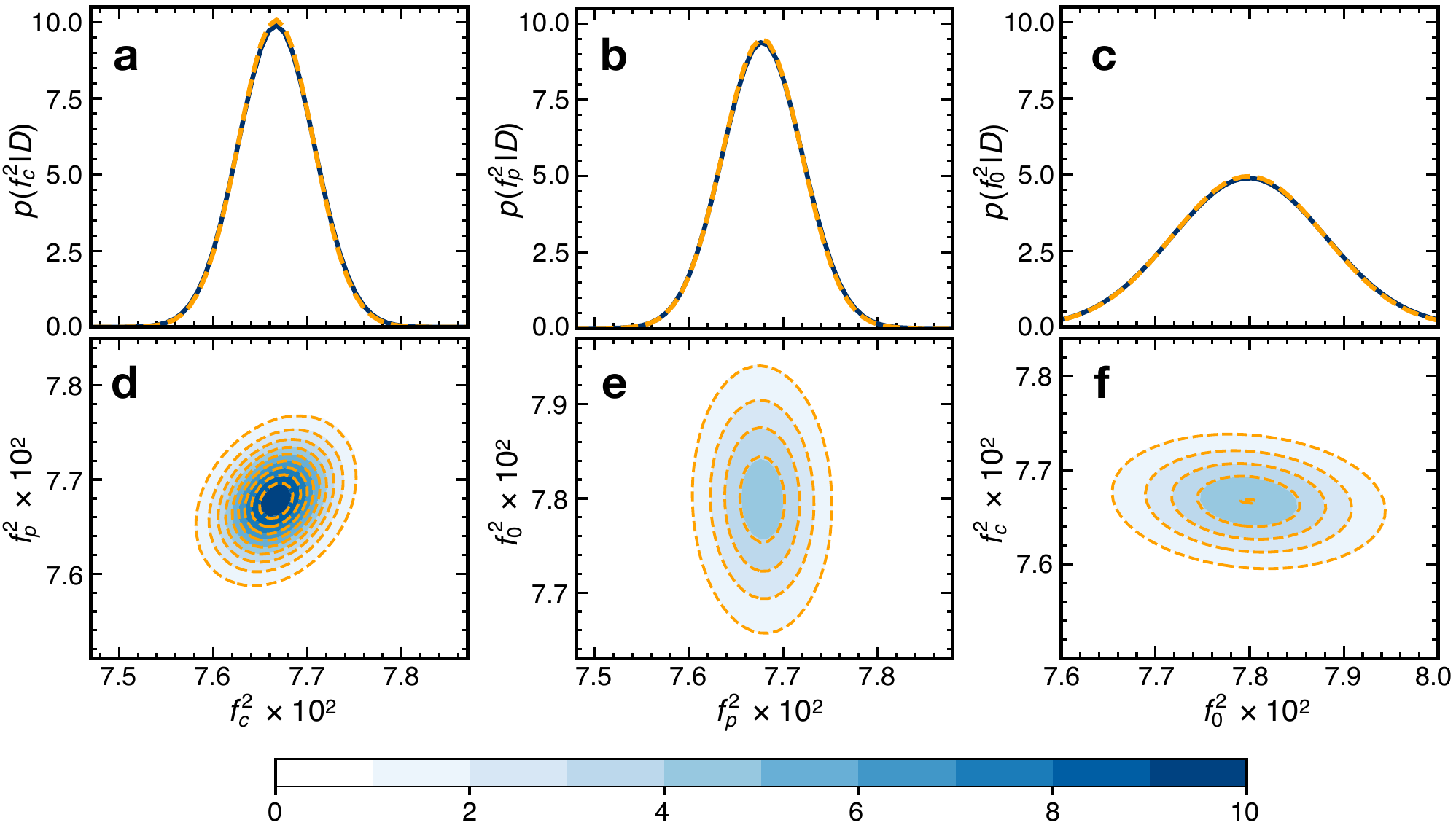}  
  \end{center}
  \vskip -0.4 true cm
  \caption{Marginal posteriors 
    for the central model and the energy range of $E_{\rm lab} = 0-280$~MeV.  
  {\bf a},  {\bf b} and  {\bf c} show the probability distributions $p(f_i^2 \vert D)$  in units of $10^2$.  
   {\bf d},  {\bf e} and  {\bf f} show the joint distributions $p({f_i^2,f_j^2} \vert D)$ 
  in units of $10^5$. Blue solid lines and filled contours are based on the 
  exact numerical evaluation,
  %and subsequent marginalization of
%  Eq.~\eqref{eq:pdf-final} of Methods, 
  while orange dashed lines/contours represent its approximation by a multivariate 
  Gaussian distribution as described in the text.
  \label{fig2}
}
\end{figure}
Although no further assumptions regarding the shape of the
distributions $p(f^2 \vert D)$ have been made, the calculated PDFs
$p(f^2 \vert D)$ are found to follow a
multivariate Gaussian distribution to a very high accuracy.
This is exemplified in Fig.~\ref{fig2} for the case of the 
central model and the energy range of $E_{\rm lab} = 0-280$~MeV.  
The distributions $p(f^2 \vert D)$ can, therefore, be accurately
characterized by the central values and errors
of the $f_i^2$'s along with the
corresponding correlation coefficients, which 
greatly facilitates their averaging as explained in Sec.~6 
of the Supplementary Material. 

We can verify the statistical validity of our results by
studying the likelihood at the optimal values of the
parameters. For example, fitting the LECs $C$ for
%the optimal value
$\Lambda = 463.5$~MeV (see Fig.~1 of Supplemented Material) and the central values of the $f^2$'s
from Eq.~(\ref{FinalResult}) in the range of $E_{\rm lab} = 0-280$~MeV
yields $\chi^2 = 4950.72$ for $N_{\rm dat} = 4926$, leading to 
$\chi^2/N_{\rm dat} =1.005$. 
The quantity $\chi^2/ (N_{\rm dat} -
N_{\rm par}) - 1= 0.012$, where $N_{\rm par} = 34$ (including $\Lambda$), is comparable to
half of the standard
deviation (s.d.), $\sqrt{2/ (N_{\rm dat} - N_{\rm par})} = 0.020$,
expected for a perfect model. 

We have also investigated  the robustness of our results with respect
to the variation of input parameters. 
Specifically, the
uncertainty from higher-order $\pi$N LECs entering the
two-pion-exchange potential is quantified by repeating our analysis
for $50$ sets of these LECs generated from the central values and
covariance matrix of
Ref.~\cite{Hoferichter:2015tha}. Furthermore, the uncertainty of the QCD
contribution $\delta m^{\rm QCD}$, even taking its conservative
estimate  of $\pm 0.30$~MeV \cite{Gasser:1974wd}, is found to induce
errors in $f_i^2$ that are negligibly small compared to the ones
given below. 
Our final result for the $\pi$N coupling constants after the
% Bayesian
averaging reads
\begin{eqnarray}
  \label{FinalResult}
    \fp^2 &=& 0.0770(5)^{\rm a}(0.8)^{\rm b}\,, \nn
              f_0^2 &=&  0.0779(9)^{\rm a} (1.3)^{\rm b} \,,\nn
                        \fc^2 &=& 0.0769(5)^{\rm a} (0.9)^{\rm b}  \,,
\end{eqnarray}
where the first error ($\rm a$) is obtained from the marginal
posteriors $p(f^2 \vert D)$ and includes the 
statistical and 
systematic errors due to the truncation of the EFT
expansion, the choice of the energy range and the associated 
data selection. The second error  ($\rm b$) reflects the uncertainty in the
higher-order $\pi$N LECs. Notice that the present study can,
  in principle, be extended to obtain a joint 
posterior probability distribution for  $f_i$'s and the higher-order
$\pi$N  LECs that can be useful for uncertainty quantification in
chiral EFT via a combined analysis of the NN data and the
experimental/empirical information on the $\pi$N scattering
amplitude, see Ref.~\cite{Carlsson:2015vda} for a related work.

\begin{figure}
  \begin{center}
\includegraphics[width=0.49\textwidth,keepaspectratio,angle=0,clip]{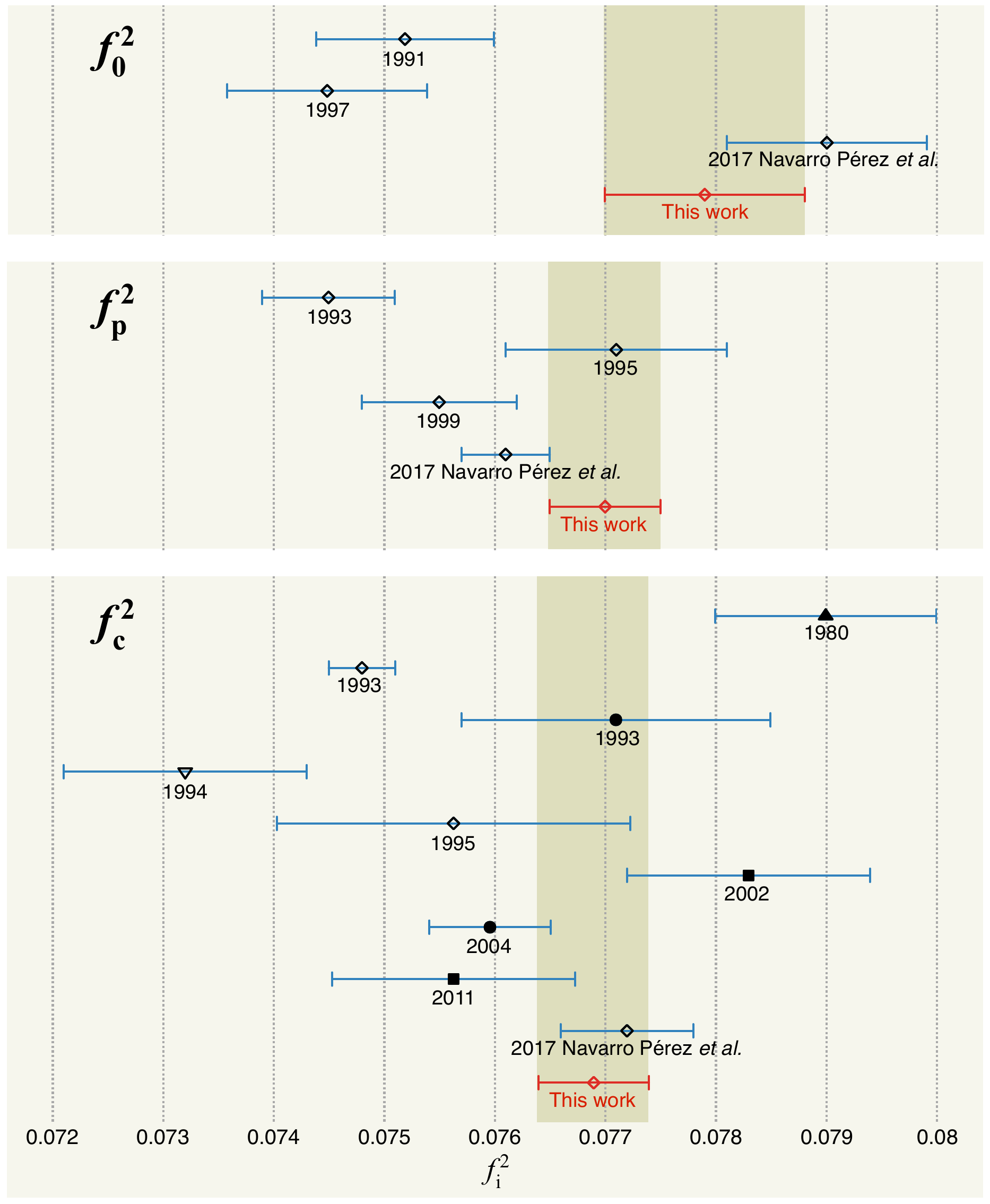}  
\end{center}
\vskip - 0.4 true cm
\caption{Values for the $\pi$N
    coupling constants. The data points show selected determinations
  of the $\pi$N coupling constants $f_0^2$, $\fp^2$ and $\fc^2$. The
  results were obtained using $\pi$N PWA \cite{Koch:1980ay} (filled triangle),
  fixed-$t$ dispersion relations of $\pi$N
scattering \cite{MarkopoulouKalamara:1993rv,Arndt:2003if}  (filled dots), $\pi$N scattering lengths in
combination with 
the GMO sum rule \cite{Ericson:2000md,Baru:2011bw} (filled squares),
proton-antiproton PWA \cite{Timmermans:1994pg} (open
triangle) and NN PWA \cite{Klomp:1991vz,Stoks:1992ja,Bugg:1994ph,deSwart:1997ep,Rentmeester:1999vw} including the 2017 Granada PWA from
Navarro P\'{e}rez {\it et al.} \cite{Perez:2016aol} 
(open diamonds). 
When provided separately, the statistical
and systematic uncertainties are added in quadrature. The vertical
bands show our full uncertainty. Uncertainties are one s.d.
\label{fig3}
}
\end{figure}

{\it Discussion of the results.}---Our results for $f_i^2$ are compared in Fig.~\ref{fig3} with selected
earlier determinations.
Similarly to the Granada PWA, we find
  considerably larger values for the coupling constants as compared to
  the ones recommended by the Nijmegen group \cite{Stoks:1992ja}. As already found
in Ref.~\cite{Perez:2016aol}, this difference cannot be
explained by the new experimental data since 1993 (see Supplemental
Material for more details), thus pointing towards
a possible sizeable systematic uncertainty from the interaction
modeling
in the Nijmegen PWA.
%Our value for $\fc^2$ is consistent with all
%quoted determinations from the $\pi$N system (at the $1.3\sigma$
%level).
Our value for $\fc^2$ is consistent with the
determinations from the $\pi$N system  in Refs.~\cite{MarkopoulouKalamara:1993rv,Arndt:2003if,Ericson:2000md,Baru:2011bw} (at the $1.3\sigma$
level).
The results for $f_0^2$
and  $\fc^2$ agree within errors with the recent determination by
the Granada group \cite{Perez:2016aol}, while for $\fp^2$ we obtain a slightly
larger value. 
However, contrary to the Granada group that found
evidence that the coupling of neutral pions to neutrons is
larger than to protons, $f_0^2-\fp^2=0.0029(10)$ \cite{Perez:2016aol}, our result
$f_0^2-\fp^2=0.0010(10)^{\rm a}(2)^{\rm b}$ is consistent with no charge dependence.
% of the  $\pi$N coupling constants.
This difference may point to significant systematic uncertainties in
the analysis by the Granada group \cite{Perez:2016aol} which are not quantified in
that paper, in particular due to the cutoff radius $r_c$
and phenomenological modeling of the interaction. In contrast,
our analysis relies on the systematically improvable EFT framework and
takes into account model-independent long- and intermediate-range
nuclear interactions due to exchange of virtual pions and photons. 
This allows us to substantially reduce the number of adjustable
parameters ($33$ in our analysis versus $55$ in
Ref.~\cite{Perez:2016aol}) while still achieving at least a comparable
description of NN data below the pion production threshold.
Compared to the Granada analysis, we also do not find a large 
anticorrelation between $f_0^2$ and $\fc^2$, see Table V of
Supplemental Material. 

In summary, our Bayesian determination of the 
$\pi$N coupling constants from np and pp scattering data 
in the framework of chiral EFT yields new reference values for these 
fundamental observables, accurate at the percent level.
It provides new insights into the isospin symmetry of the strong
interaction at the hadronic level by quantifying the charge dependence
of these quantities. Our work also establishes important benchmarks for future first
principles calculations using lattice QCD and QED, see \cite{Borsanyi:2014jba} for a
first step along this line, and opens the door for
precision studies of nuclear structure and reactions by fixing
the strengths of the long-range nuclear forces.
As a very recent example, we mention the high-accuracy
calculation of the deuteron charge and quadrupole form factors
\cite{Filin:2020tcs}, where the isospin-breaking corrections considered in
this work have been taken into account and were found to play
  an important role for the determination of the quadrupole moment,
  for which the value of 
  $Q_d=0.2854\substack{+0.0038\\ -0.0017}\ \text{fm}^2$ was obtained.
  %In particular,
  Redoing the same analysis with the IB corrections from this paper
  being 
  switched off, i.e.~using the NN interactions from \cite{Reinert:2017usi}, the
  central value of the quadrupole moment would change significantly to $Q_d=0.2803 \
  \text{fm}^2$. These calculations are currently being extended to
  other light nuclei, and we expect the considered IB corrections to
  be relevant  at the desired accuracy level. It would also be
  interesting to explore the implications of our study for the 
  understanding of charge symmetry
  breaking in binding energy differences of mirror nuclei \cite{Friar:2004rg} and
 for  certain low-energy three-nucleon scattering observables such as the
  doublet scattering length \cite{Witaa:2003en} and vector analyzing power $A_y$
  \cite{Witala:1994zz}.
  Furthermore, since our analysis also results in an accurate determination of the IB
  short-range operators in the $^1$S$_0$ partial wave, it may
  shed new light on the ongoing studies of neutrinoless  double-beta decay
  in chiral EFT, see Ref.~\cite{Cirigliano:2019vdj} for a related discussion. This will
  be addressed in a separate publication.
Last but not least, we emphasize that the value for $f_c$ can be related
to the pionic hydrogen width $\Gamma_{1s}$, whose measurement  at
PSI is currently being analyzed, see \cite{Gotta:2015oza} for a preliminary
result. 

\bigskip
 We are grateful to Martin Hoferichter, Ulf-G. Mei{\ss}ner and Enrique Ruiz Arriola for sharing their insights into
   the considered topics and useful discussions. We also thank Ulf-G. Mei{\ss}ner for helpful comments on the manuscript. This work was
   supported by BMBF (contract No. 05P18PCFP1) and by DFG
   through funds provided to the Sino-German CRC 110 ``Symmetries and
   the Emergence of Structure in QCD'' (Grant No. TRR110).

%\putbib[piNPRL_sorted]
%\end{bibunit}{}
%Control: production of eprint (1) enabled
%

\makeatother

\beginsupplement
\onecolumngrid

%\begin{bibunit}[apsrev4-2]
%\nocite{apsrev42Control}

\newpage

\section{Supplemental Material}

\subsection{1.~Chiral effective field theory for the NN potential}

The functional form of the NN interaction employed in our analysis
corresponds to the N$^4$LO$^+$ potential of Ref.~\cite{Reinert:2017usi_suppl}, supplemented with
the CIB and CSB interactions up to N$^4$LO as described below.
\begin{itemize}
\item  
We employ the most general form of the static OPE
potential given by 
\begin{equation}
  V_{1\pi}^{\rm pp} = \fp^2 \; V(M_{\pi^0}), \; \quad \quad
  V_{1\pi}^{\rm nn} = \fn^2 \; V(M_{\pi^0}), \; \quad \quad
V_{1\pi}^{\rm np} = -f_0^2 \; V(M_{\pi^0}) + (-1)^{I+1} 2 \fc^2 \; V(M_{\pi^\pm}), 
\end{equation}
where $I=0,1$ is the isospin quantum number of the np
system and $V (M_i)$ is given by 
\begin{equation}
V(M_i) = -\frac{4\pi}{M_{\pi^\pm}^2} \frac{\vec \sigma_1  \cdot \vec q
  \, \vec \sigma_2 \cdot \vec q}{\vec q^{\; 2}+M_i^2}\,.
\end{equation}
Here, $\vec \sigma$ are Pauli spin matrices and $\vec q = \vec p\,' -
\vec p$ is the momentum transfer of the nucleons, with $\vec p \, '$
and $\vec p$ denoting the final and inititial momenta. 
\item
Electromagnetic corrections to the OPE start to contribute at
N$^3$LO. The expressions for the corresponding $\pi \gamma$-exchange
potential at this order have been derived
%using dimensional
%regularisation
in Ref.~\cite{vanKolck:1997fu_suppl} and depend only on the
known quantities $\alpha$, $F_\pi$, $g_A$ and the pion masses. 
We employ the {\it convention} to absorb
the pion-pole contribution to the $\pi \gamma$-exchange potential into
a definition of $\fc^2$.
\item
At the order considered, the IB two-pion-exchange potential involves
the following structures
\begin{equation}
V_{2 \pi} = \big(V_C + V_S  \, \vec \sigma_1 \cdot \vec \sigma_2 + V_T
\, \vec
\sigma_1 \cdot \vec q \, \vec \sigma_2 \cdot \vec q \, \big) \, \tau_1^3 \tau_2^3 
+
              \big(W_C + W_S  \, \vec \sigma_1 \cdot \vec \sigma_2   
  + W_T
\, \vec
              \sigma_1 \cdot \vec q \, \vec \sigma_2 \cdot \vec q \, \big)
 \big( \tau_1^3 + \tau_2^3 \big)\,,
\end{equation}
with the first and second lines showing the CIB and CSB operators,
respectively. The leading IB TPE contributions appear at fourth order (N$^3$LO).
The corresponding unregularized expressions for $V_C^{(4)}$ and  $W_{C,S,T}^{(4)}$
% obtained using dimensional regularisation,
are given in Eqs.~(3.40) and (3.47) of Ref.~\cite{Epelbaum:2005fd_suppl}
with $\Lambda = \infty$, while $V_{T}^{(4)} = V_{S}^{(4)} = 0$.  
% No contributions to $V_{T,S}$ appear at this order.
The expressions
at fifth order (N$^4$LO) are given in Eqs.~(3.49), (3.52)
of Ref.~\cite{Epelbaum:2005fd_suppl} for $V_{T,S}^{(5)}$ and in Eq.~(2.11) of 
Ref.~\cite{Epelbaum:2008td} for $W_{C}^{(5)}$.  For the remaining 
IB TPE contributions at this order, we employ the expressions
\begin{eqnarray}
    W^{(5)}_T(q) &=& -\frac{1}{q^2} W^{(5)}_S(q) =  - \bigg(\frac{g_A^2}{16\pi^2F_\pi^4}\delta m \, c_4 
    + \frac{4 \fc^3}{M_{\pi^\pm}^4}(\fp-\fn)\bigg) L(q),
    \nonumber\\
    V^{(5)}_C(q) &=& -\frac{\fc \, (2 \fc - \fp - \fn) }{24 \pi F_\pi^2
                     M_{\pi^\pm}^4} \bigg(\frac{16 \pi F_\pi^2 \fc^2
                     }{4 M_\pi^2 + q^2}
                     (128 M_\pi^4 + 112 M_\pi^2 q^2 + 23 q^4)
    - (8 M_\pi^2 + 5 q^2) M_{\pi^\pm}^2\bigg) L(q)\,,
\end{eqnarray}
where $q \equiv |
\vec q \,|$,
$M_\pi = 2/3 \, M_{\pi^\pm} + 1/3 \, M_{\pi^0}$ and the function
$L(q)$ is given by $L(q) = s/q \ln ((s+q)/(2 M_\pi)$ with $s =
\sqrt{q^2 + 4 M_\pi^2}$. The above expressions generalize the ones
of Ref.~\cite{Epelbaum:2005fd_suppl} by including all TPE terms
linear in the differences between $f_i$'s with no assumptions about 
the dominance of CSB over CIB.

The employed  IB TPE depends on the known
LECs $F_\pi$, $g_A$, $c_1$, $c_2$, $c_3$, $c_4$, pion masses, the
neutron-proton mass differences $\delta m$ and $\delta m^{\rm QCD}$
and the IB $\pi$N coupling constants $\fp$, $\fn$ and $\fc$ to be determined.  
\item
Up to the considered order, the CIB and CSB 
  short-range interactions contributing to the np and pp systems
  involve two terms in the $^1$S$_0$ channel and one term in each of the
  $^3$P$_0$, $^3$P$_1$ and $^3$P$_2$ partial waves, see Ref.~\cite{Epelbaum:2005fd_suppl}
  for explicit expressions. The corresponding
  LECs have to be extracted from NN data. 
\end{itemize}
Regularization of the IB potential is performed
in the same way as in Ref.~\cite{Reinert:2017usi_suppl}. In particular,
contact interactions are multiplied with a nonlocal
Gaussian cutoff, while regularization of the long-range contributions is carried out 
by utilizing the spectral representation. For the contributions to
the $\pi \gamma$-exchange potential and $V_C^{(4)}$ that feature a pole at the branch
point, the spectral representation is employed for the corresponding
indefinite integral. 

Having defined the functional form of the employed NN potential, we
are now in the position to specify the numerical values of various
parameters. The long-range interactions stemming from the
pion, pion-photon and two-pion exchanges are completely determined by
the $\pi$N coupling constants $\fp^2$, $f_0^2$ and
$\fc^2$ treated as free parameters, the cutoff value $\Lambda$ to be
marginalized over and the known masses and constants 
specified in Table \ref{tab1}.
To account for the Goldberger-Treiman
discrepancy, we replace $g_A$ in all expressions
%for the
%pion-photon and
%two-pion exchange
%potential
with a rescaled effective  axial vector coupling
$g_A^{\rm eff}$.  For consistency reasons, we employ the 
values for $F_\pi$ and $g_A^{\rm eff}$ from the determination of the $\pi$N LECs in
Ref.~\cite{Hoferichter:2015tha_suppl}.  
For the QCD contribution to the nucleon mass difference, we use the
value $\delta m^{\rm QCD} = 2.05(30)$~MeV \cite{Gasser:1974wd_suppl} which is
compatible with both the lattice QCD and QED result of
Ref.~\cite{Borsanyi:2014jba_suppl} obtained by using the experimental value
of  $\delta m$, $\delta m^{\rm QCD} \simeq 2.16$~MeV, and with the
recent determination by the same authors $\delta m^{\rm
  QCD} \simeq 1.87(16)$~MeV \cite{Gasser:2020mzy_suppl}.
The short-range part of the
% N$^4$LO$^+$ NN
potential involves $25$
isospin-invariant contact interactions, see Eqs.~(A.1) and (17) of
Ref.~\cite{Reinert:2017usi_suppl}, and $5$ IB contact terms as explained above.  
The strengths $C_i (\Lambda )$ of the corresponding interactions are treated as
free parameters.

\begin{table}[t]
\caption{Parameters for the long-range part of the NN potential  \label{tab1}}
\smallskip
\begin{tabular*}{\textwidth}{@{\extracolsep{\fill}}cr}
   \noalign{\smallskip}
\hline 
  \noalign{\smallskip}
  Quantity & Employed value \\
  \noalign{\smallskip}
  \hline
  \noalign{\smallskip}
  Masses and mass differences: & \\[1pt]
  $M_{\pi^\pm}$ & $139.57$ MeV \\[1pt] 
  $M_{\pi^0}$ & $134.98$ MeV \\[1pt]
  $m_p$ & $938.272$ MeV \\[1pt] 
  $m_n$ & $939.565$ MeV \\[1pt]
  $\delta m^{\rm QCD}$, Ref.~\cite{Gasser:1974wd_suppl} & $2.05(30)$ MeV\\
   \noalign{\smallskip}
  \hline
   \noalign{\smallskip}
  Fine-structure constant
  $\alpha$: & $1/137.036$  \\
 \noalign{\smallskip}
  \hline
   \noalign{\smallskip}
  Pion decay constant
  $F_\pi$: & $92.2$ MeV  \\
 \noalign{\smallskip}
  \hline
   \noalign{\smallskip}
  Effective axial-vector coupling of the nucleon
  $g_A^{\rm eff}$: & $1.289$ \\
 \noalign{\smallskip}
  \hline
   \noalign{\smallskip}
  Pion-nucleon LECs from Ref.~\cite{Hoferichter:2015tha_suppl}: &  \\[1pt]
  $c_1$ & $-1.10(3)$ GeV$^{-1}$  \\[1pt]
  $c_2$ & $3.57(4)$ GeV$^{-1}$  \\[1pt]
  $c_3$ & $-5.54(6)$ GeV$^{-1}$  \\[1pt]
  $c_4$ & $4.17(4)$ GeV$^{-1}$  \\[1pt]
  $\bar d_1 + \bar d_2$ & $6.18(8)$ GeV$^{-2}$  \\[1pt]
  $\bar d_3$ & $-8.91(9)$ GeV$^{-2}$  \\[1pt]
  $\bar d_5$ & $0.86(5)$ GeV$^{-2}$  \\[1pt]
  $\bar d_{14} - \bar d_{15}$ & $-12.18(12)$ GeV$^{-2}$  \\[1pt]
  $\bar e_{14}$ & $1.18(4)$ GeV$^{-3}$  \\[1pt]
  $\bar e_{17}$ & $-0.18(6)$ GeV$^{-3}$  \\
 \noalign{\smallskip}
  \hline
\end{tabular*}
\end{table}

The NN potential specified above defines our central model (Model 1)
used in the
% Bayesian
determination of $f^2$. To estimate the
systematic uncertainty from the truncation of the EFT expansion of the
IB NN potential, we also consider two additional interaction models that include
selected IB contributions at sixth order (i.e.~N$^5$LO):
\begin{itemize}
\item[Model 2:] Same as Model 1 but including the subleading
  $\pi \gamma$-exchange potential driven by the numerically large
  isovector magnetic moment $\kappa_v \simeq 4.706$ of the nucleon,
  which has been  worked out
  in Ref.~\cite{Kaiser:2006ws}.   
\item[Model 3:] Same as Model 1 but including the IB TPE potential
  $\propto  (f_j - f_k) \, c_i$. These contributions have not been considered
  before. The unregularized expressions read:
  \begin{eqnarray}
    V^{(6)}_T(q) &=& -\frac{1}{q^2} V^{(6)}_S(q) =
                     -\frac{\fc (2 \fc-\fp-\fn) c_4}{4F_\pi^2
                     M_{\pi^\pm}^2} \, \big(4M_\pi^2+q^2 \big) \, A(q) \,, 
    \nonumber\\
    W^{(6)}_C(q) &=& -\frac{\fc (\fp-\fn)}{2 F_\pi^2 M_{\pi^\pm}^2}
                     \, \big(2M_\pi^2+q^2\big)\, \big(4M_\pi^2
                     c_1-c_3(2M_\pi^2+q^2)\big) \, A(q) \,,
  \end{eqnarray} 
where the function $A (q)$ is given by $A(q)=1/(2q) \arctan (q/(2 M_\pi))$.  
\end{itemize} 

\subsection{2.~Calculation of the scattering amplitude in the presence of
electromagnetic interactions}
In addition to the finite-range NN force described above, it is
necessary to take into account the long-range electromagnetic
interactions when calculating the scattering amplitude. Throughout
this work, we include the so-called improved Coulomb
potential \cite{Austen:1983te},
the magnetic moment interaction \cite{Stoks:1990us} as well as the
vacuum-polarization potential \cite{Durand:1957zz} as described in detail in
Ref.~\cite{Reinert:2017usi_suppl}. These effects have also been taken into account
in the PWAs by the Nijmegen \cite{Stoks:1993tb_suppl} and Granada \cite{Perez:2013jpa_suppl,Perez:2016aol_suppl} groups. 
The nuclear amplitude is calculated by solving the Lippmann-Schwinger
equation in the partial wave basis including all channels up to $j=20$.

\subsection{3.~Objective function}
To determine the values of $\fp^2$, $f_0^2$ and $\fc^2$ from np and pp scattering
data, we employ the same augmented $\chi^2$ measure as in
Ref.~\cite{Reinert:2017usi_suppl}.
Specifically, for the $j^{\rm th}$ experiment consisting of $n_j$ measurements
$\{O_{j,i}^{\rm exp}\}_{i=1,...,n_j}$ of some observable and their corresponding
statistical
% measurement
errors $\{\delta O_{j,i}^{\rm exp}\}_{i=1,...,n_j}$, we define
its $\chi^2$ measure as 
\begin{equation}  
    \label{eq:chi2-dataset}
    \chi^2_j = \sum_{i=1}^{n_j} \left(\frac{O_{j,i}^{\rm exp}-Z_jO_{j,i}^{\rm
        theo}}{\delta O_{j,i}}\right)^2 + \left( \frac{Z_j-1}{\delta_{\rm
        norm, j}}\right)^2.
\end{equation}
Here, $\{O_{j,i}^{\rm theo}\}_{i=1,...,n_j}$ are the theoretical values of the
observable calculated from the NN potential and $\delta_{\rm
  norm,j}$ is the (systematic) normalization error of the dataset $j$. If $\delta_{\rm
  norm,j} \ne 0$, the optimal norm is estimated by minimizing
$\chi^2_j$ with respect to the factor $Z_j$, see Ref.~\cite{Reinert:2017usi_suppl}
for further information. Our total $\chi^2$ is the sum of the
contributions of
np and pp scattering data in a given energy range and two
additional data points for the deuteron binding energy $B_d$ and the
np coherent scattering length $b_{\rm np}$: 
\begin{equation}
    \label{eq:chi2}
    \chi^2 = \sum_j \chi^2_j + \left(\frac{B_d-B_d^{\rm exp}}{\delta B_d}\right)^2 + \left(\frac{b_{\rm np}-b_{\rm np}^{\rm exp}}{\delta b_{\rm np}}\right)^2,
\end{equation}
with $B_d^{\rm exp} = 2.224575(9)$~MeV \cite{VanDerLeun:1982bhg} and $b_{\rm np}^{\rm exp} =
-3.7405(9)$~fm \cite{Schoen:2003my}. Following Ref.~\cite{Reinert:2017usi_suppl}, we use a slightly
relaxed error for $B_d$, $\delta B_d = 5 \times 10^{-5}$~MeV, and employ the 
experimental uncertainty of $b_{\rm np}$ for $\delta b_{\rm np}$.

\subsection{4.~Data selection} Following the same criteria as 
Refs.~\cite{Stoks:1993tb_suppl,Gross:2008ps,Perez:2013jpa_suppl}, we only consider 
NN data published in peer-reviewed journals and do not 
take into account pp total cross sections or quasi-elastic scattering 
data, i.e.~data that have been extracted from three-body experiments.
It is well known that not all available NN scattering
data are mutually compatible within statistical errors. To determine
mutually compatible np and pp data, we follow the standard iterative 
procedure to reject $3\sigma$-inconsistent data \cite{Gross:2008ps,Perez:2013jpa_suppl,Perez:2016aol_suppl}. 
Specifically, we start from a fit to all considered NN data which yields 
$\chi^2$/datum $> 1$. If the residuals of the objective function are properly 
normal-distributed, then $\chi^2_j$ of Eq.~\eqref{eq:chi2-dataset} should 
follow a $\chi^2$-distribution with $n_j$ degrees-of-freedom ($n_j-1$ for 
floated datasets, see Ref.~\cite{Reinert:2017usi_suppl}). If for some dataset $\chi^2_j$ 
is either too high or too low, i.e. if the probability of obtaining that 
value is $\le 0.27 \%$, the dataset is rejected. The potential is 
refitted to the accepted data and the compatibility of \emph{all} data is tested 
again, resulting in a new set of accepted datasets. This process is then repeated 
until the data selection has reached stability.

For the selection process, simultaneous fits of all contact LECs and the $\pi$N 
constants $\fp^2$, $f_0^2$ and $\fc^2$ were performed using the central model (Model 1).
We used a fixed cutoff value of $\Lambda = 450$~MeV which was shown in Ref.~\cite{Reinert:2017usi_suppl} 
to yield the best description of NN data among the considered cutoffs and 
whose choice is verified a posteriori by being close to the optimal  
value $\Lambda = 463.5$~MeV of this analysis. For each value of $E_\mathrm{lab}^\mathrm{max}$ 
employed in this work, we have performed a separate data selection,
and we found 
consistent results for $E_\mathrm{lab}^\mathrm{max}=260-300$~MeV. Our data selection is largely 
in agreement with the Granada 2013 database from Ref.~\cite{Perez:2013jpa_suppl} and, 
for the sake of brevity, we restrict ourselves to listing differences between our 
data selection at $E_\mathrm{lab}^\mathrm{max} =300$~MeV and the Granada 2013 database in 
Tables~\ref{tab:np-data-differences} and \ref{tab:pp-data-differences}.
As mentioned above, both tables also give our database for
$E_\mathrm{lab}^\mathrm{max} =260$ 
and $280$~MeV if restricted to the corresponding energy ranges. For pp data with 
$E_\mathrm{lab}^\mathrm{max} \le 240$~MeV, the 3 differential cross sections of PA(58) get rejected, while 
the 15 differential cross sections of PA(58) and CO(67) rejected at higher energies 
are compatible with the rest of the data based on the $3\sigma$ criterion. The CO(67) dataset, in particular, consists of 
measurements of remarkable precision that are sensitive to F-waves, see Ref.~\cite{Reinert:2017usi_suppl} 
for details. While it is properly described by our potential within estimated truncation errors 
for all $E_\mathrm{lab}^\mathrm{max}$, its $\chi^2$ value is slightly too high to be statistically 
accepted for $E_\mathrm{lab}^\mathrm{max} \ge 260$~MeV. For np data with $E_\mathrm{lab}^\mathrm{max} \le 240$~MeV,
KA(63) gets rejected, and the dataset of Ref.~\cite{MEASDAY1966142} is
accepted for $E_\mathrm{lab}^\mathrm{max} =240$~MeV only.

\begin{table}[t]
  \caption{Differences in the np data selection of this work with
      $E_\mathrm{max}=300$~MeV in comparison to the Granada 2013
      database \cite{Perez:2013jpa_suppl}
      \label{tab:np-data-differences}}
    \smallskip
    \begin{tabular*}{\textwidth}{@{\extracolsep{\fill}}lllrccc}
        %\toprule
 \noalign{\smallskip}
      \hline
       \noalign{\smallskip}
        $E_\textrm{lab}$ [MeV] & Ref. Code &  Obs.      & $n$ &  comment & Ref. \\
        %\midrule
 \noalign{\smallskip}
      \hline
       \noalign{\smallskip}
        \multicolumn{6}{l}{additionally included data} \\[1pt]
             0.509-2.003 & PO(82) & $\sigma_\textrm{tot}$  &  3 &     & \cite{Poenitz:1982qdn}  \\[1pt] 
    1.005-2.530 & FI(54) & $\sigma_\textrm{tot}$  &  2 &                          & \cite{Fischer:1954}     \\[1pt]
          2.535 & DV(71) & $\sigma_\textrm{tot}$  &  1 &                          & \cite{Davis:1971zza}    \\[1pt]
           25.8 & OC(91) & $D_t$                  &  1 & (a)       & \cite{Ockenfels:1991qao}\\[1pt]
           77.0 & WH(60) & $P$                    &  8 & (a)       & \cite{Whitehead:1960}   \\[1pt]
           77.0 & WH(60) & $P$                    &  9 & (a)       & \cite{Whitehead:1960}   \\[1pt]
          199.0 & TH(68) & $P$                    &  8 &                          & \cite{Thomas:1968zza}   \\[1pt] 
          200.0 & KA(63) & $d\sigma/d\Omega$      & 19 & (b)   & \cite{Kazarinov:1963}   \\[1pt] 
          300.0 & DE(54) & $d\sigma/d\Omega$      & 15 &                          & \cite{DePangher:1954}   \\
 \noalign{\smallskip}
      \hline
       \noalign{\smallskip}
        \multicolumn{6}{l}{additionally rejected data} \\[1pt]
              0 & LO(74) & $\sigma_\textrm{tot}$  &  1 & (c) & \cite{Lomon:1974zza}    \\[1pt] 
    0.155-0.795 & DA(13) & $\sigma_\textrm{tot}$  & 65 &                          & \cite{Daub:2012qb}      \\[1pt] 
           22.5 & FL(62) & $\sigma_\textrm{tot}$  &  1 & (d)  & \cite{Flynn:1962zz}     \\[1pt] 
           50.0 & FI(80) & $P$                    &  4 &                          & \cite{Fitzgerald:1980zz}\\[1pt]  
          135.0 & LE(63) & $A$                    &  5 & (e)  & \cite{Lefrancois:1963zz}\\[1pt]
          137.0 & LE(63) & $R$                    &  5 & (e)  & \cite{Lefrancois:1963zz}\\[1pt]
          197.0 & SP(67) & $D_t$                  &  3 & (e)  & \cite{Spalding:1967}    \\[1pt]
          295.0 & GR(82) & $\sigma_\textrm{tot}$, $\Delta\sigma_\textrm{T}$, $\Delta\sigma_\textrm{L}$  &  3 & (f) & \cite{Grein:1981hn}     \\
        %\bottomrule
 \noalign{\smallskip}
      \hline
    \end{tabular*}
    \begin{footnotesize}
    \begin{itemize}
        \item[(a)] data not considered in Granada database selection process
        \item[(b)] the outlier at 97.0$^\circ$ has been removed due to too high individual $\chi^2$%, in agreement with NN-Online~\cite{nn-online}
        \item[(c)] LO(74) cites the data of Ref.~\cite{Houk:1971uq}, which is already included in the database
        \item[(d)]  the original publication explicitly mentions that no total cross section has been measured
        \item[(e)]  data from quasi-elastic scattering
        \item[(f)]  dispersion relation prediction
    \end{itemize}
    \end{footnotesize}
\end{table}

\begin{table}[t]
 \caption{Same as Table~\ref{tab:np-data-differences} but for pp data
      \label{tab:pp-data-differences}}
    \smallskip
    \begin{tabular*}{\textwidth}{@{\extracolsep{\fill}}lllrcccc}
        %\toprule   
 \noalign{\smallskip}
      \hline  
 \noalign{\smallskip}
      $E_\textrm{lab}$ [MeV] & Ref. Code &  Obs.      & $n$ & comment & Ref. \\
        %\midrule
 \noalign{\smallskip}
      \hline
       \noalign{\smallskip}
        \multicolumn{6}{l}{additionally included data} \\[1pt]
          137.0 & PA(58) & $d\sigma/d\Omega$      &  3 &                         & \cite{PALMIERI1958299}  \\[1pt]
          239.9 & AL(04) & $d\sigma/d\Omega$      & 17 & (a)      & \cite{Albers:2004iw}    \\[1pt]
          255.2 & AL(04) & $d\sigma/d\Omega$      & 18 & (a)      & \cite{Albers:2004iw}    \\[1pt]
          270.8 & AL(04) & $d\sigma/d\Omega$      & 19 & (a)      & \cite{Albers:2004iw}    \\[1pt]
          286.7 & AL(04) & $d\sigma/d\Omega$      & 20 & (a)      & \cite{Albers:2004iw}    \\
 \noalign{\smallskip}
      \hline
       \noalign{\smallskip}
        \multicolumn{6}{l}{additionally rejected data} \\[1pt]
          144.0 & JA(71) & $d\sigma/d\Omega$      & 27 &                         & \cite{Jarvis:1971fla}   \\[1pt]
          144.1 & CO(67) & $d\sigma/d\Omega$      & 15 &                         & \cite{Cox:1968jxz}      \\[1pt]
          147.0 & PA(58) & $d\sigma/d\Omega$      & 15 &                         & \cite{PALMIERI1958299}  \\[1pt]
          217.0 & TI(61) & $P$                    &  6 & (e) & \cite{Tinlot:1961jww}   \\
        %\bottomrule
 \noalign{\smallskip}
      \hline
    \end{tabular*}
\end{table}

In addition, we have 
%performed a cleanup of datasets whose inclusion remains unchanged. In
%particular, we
corrected minor misprints in Ref.~\cite{Perez:2013jpa_suppl}  for
the data of Refs.~\cite{Flynn:1962zz,Fink:1990cmm} and removed
erroneously assigned normalization errors $\delta_{\mathrm{norm,j}}$
to the datasets of
Refs.~\cite{Ockenfels:1991qao,Amsler:1977av,Greeniaus:1979zn} which
only 
contain a single data point.

\subsection{5.~Calculation of the marginalized PDF $\mathbf{p(f^2 \vert D)}$}
In our Bayesian analysis, we employ an uninformative uniform prior for
the cutoff $\Lambda$, namely $p(\Lambda) = (\Lambda_{\rm
  max}-\Lambda_{\rm min} )^{-1}$ for $\Lambda \in [\Lambda_{\rm
  min},\Lambda_{\rm max}] $, while $0$ elsewhere.
Similarly, we use a uniform prior for the $\pi$N coupling constants
$f_i^2$, which is non-zero in a sufficiently large region $B = \left\{f^2 \, \vert \; f^2_{i,\rm min} \le f_i^2 \le f^2_{i,\rm
    max}\right\}$ around
their previously obtained values.
Finally, we employ a Gaussian prior for the LECs $C_i$,
 $p(C) = (\sqrt{2\pi} \bar{C})^{-n}
 \exp (-\vec{C}^2 /(2\bar{C}^2))$,
 with $n=30$ being the dimension of $\vec C$,
to encode the
naturalness assumptions for their values. Here, the parameter $\bar C$
to be specified below is
dimension-less, and the (dimensionful) $C_i$'s in the spectroscopic
notation are expressed in their natural units of $4 \pi/(F_\pi^2 \Lambda_{\rm
  b}^{2k})$ with $2k$ being the power of
momenta and $\Lambda_{\rm b} =
650$~MeV, see Ref.~\cite{Reinert:2017usi_suppl}. 

Using the approximation for  the likelihood $p(D \vert f^2
C\Lambda)$ in Eq.~(3) of the main text and the explicit
form of the priors $p(\Lambda)$,
$p(f^2)$ and $p(C)$, the expression for the
marginalized posterior in Eq.~(2) of the main text takes the form
\begin{eqnarray}
    \label{eq:pdf-final}
&&p(f^2 \vert D) = \frac{1}{\tilde{N}} \int_{\Lambda_{\rm min}}^{\Lambda_{\rm max}} d\Lambda \; \frac{1}{\sqrt{\det A}} e^{-\frac{1}{2}(\chi^2_{\rm min} + \frac{1}{\bar{C}^2}C_{\rm min}^TC_{\rm min} - \frac{1}{\bar{C}^4}C_{\rm min}^T A^{-1} C_{\rm min})} \, \mathbb{1}_{B}(f^2)\,,
\end{eqnarray}
where $A = \frac{1}{2}H+\frac{1}{\bar{C}^2}\mathbb{1}$, the normalization constant $\tilde{N}$ is given by  $\tilde{N} = N \bar{C}^n \, p(D) \,
(\Lambda_{\rm max}-\Lambda_{\rm min}) \prod_i(f^2_{i,\rm
  max}-f^2_{i,\rm min})$ and $\mathbb{1}_{B}$ is the indicator
function of the cuboid $B$. 

\begin{figure}[t]
  \begin{center}
\includegraphics[width=0.49\textwidth,keepaspectratio,angle=0,clip]{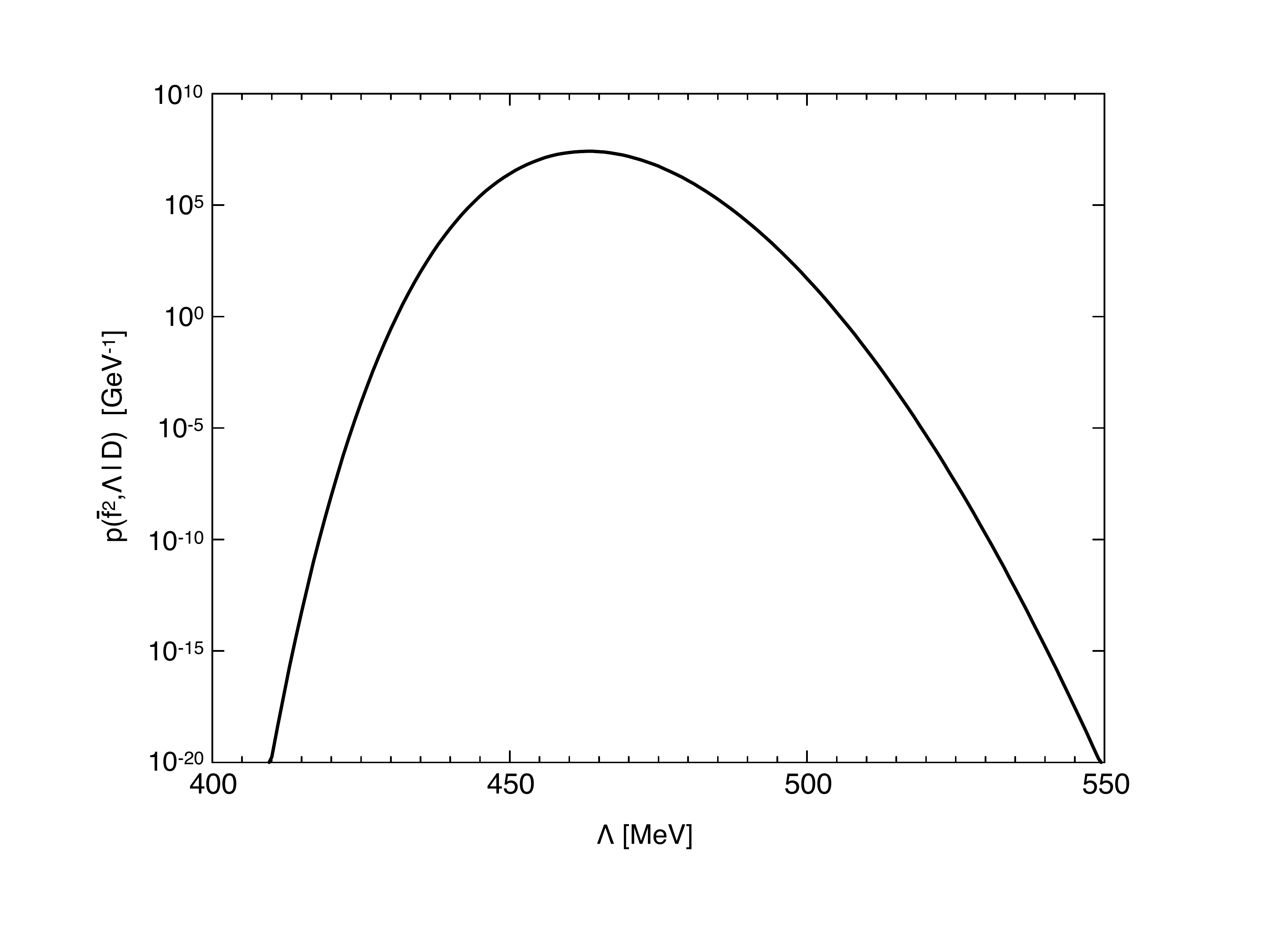}  
\end{center}
\vskip -0.4 true cm
\caption{The integrand of
    Eq.~(\ref{eq:pdf-final}) as a function of the cutoff
    $\Lambda$ for Model 1.
  $\bar f_i^2$ refer to the central values of the LECs $f_i^2$ given
  in Eq.~(4) of the main text. The data $D$ correspond to the mutually
  compatible NN scattering data according to the own selection in the range of
  $E_{\rm lab} = 0-280$~MeV. 
\label{Integrand}
}
\end{figure}

It remains to specify the parameters entering the priors $p(\Lambda)$,
$p(f^2)$ and $p(C)$. For the naturalness prior of the short-range LECs $C_i$ we choose
the value of $\bar{C} = 5$. Given the abundance of NN
scattering data, the  
$C_i$'s are well-constrained by the likelihood.  In our
previous study \cite{Reinert:2017usi_suppl}, the $C_i$'s
were found to be of natural size in the considered cutoff range of
$\Lambda$, i.e.~$|C_i | \sim 1$, even without 
imposing any naturalness constraints. In particular, none of the values of $| C_i |$
from Ref.~\cite{Reinert:2017usi_suppl} exceed $\bar{C} = 5$, see Fig.~7 of Ref.~\cite{Epelbaum:2019kcf}. We,
therefore, expect the naturalness constraint $\bar{C} = 5$ to have a
negligible effect on the obtained results. The same Gaussian prior
with $\bar{C} = 5$ was also employed in
Ref.~\cite{Wesolowski:2018lzj} in a fit to scattering phase shifts and
was reported to have only minor impact. Next, for the 
$p(\Lambda)$-prior, we employ the values of $\Lambda_{\rm min} =
400$~MeV and $\Lambda_{\rm max} = 550$~MeV. Figure~\ref{Integrand}
shows, as a representative example, the
integrand of Eq.~\eqref{eq:pdf-final} at the central values for the
$f_i^2$'s for the case of Model 1 and $E_{\rm lab} = 0-280$~MeV in the employed cutoff range.
The maximum of the integrand is at $\Lambda \simeq 463.5$~MeV, and
its contribution beyond $\Lambda = 430 - 510$~MeV has already dropped
by more than five orders of magnitude compared to its maximum and
is negligible. Thus, the cutoff is constrained by the likelihood and
not by the prior. We emphasize that the preference of
the $\Lambda = 430-510$~MeV region does not contradict the findings
of Ref.~\cite{Reinert:2017usi_suppl}, where a weak $\Lambda$-dependence of the
$\chi^2$/datum values was shown. Indeed, $\chi^2$ values have a logarithmic
scale compared to the probability density and thus vary much less.
Finally, for the uniform $p(f^2)$-prior, we choose the parameters to be
$f^2_{\rm p,  min} = f^2_{0, \rm min} = f^2_{\rm c,  min} = 0.0729$,
$f^2_{\rm p,  max} = f^2_{\rm c,  max} = 0.0812$ and $f^2_{0, \rm max} = 0.0827$. 
The chosen limits of $f_i^2$ only enter the marginal PDF, and the adequacy
of their choice becomes obvious from Fig.~2 of the main text.   
To summarize, we expect that making the priors $p(\Lambda)$,
$p(f^2)$ and $p(C)$ less informative should have a negligible effect on our results.

The remaining integration over $\Lambda$ in Eq.~(\ref{eq:pdf-final}) is performed
numerically via Gaussian quadrature. Scaling up the number of
integration points for the cutoff naively is computationally costly as
each evaluation of the integrand in Eq.~\eqref{eq:pdf-final} requires
a fit of $30$ contact LECs. However, the logarithm of the integrand is a
smooth function so that we can safely evaluate it at just $8$ points in
the range of $\Lambda = 400 - 550$ MeV and interpolate it by a polynomial.

\subsection{6.~Systematic uncertainty due to the choice of the energy
  range in the fits and neglect of IB interactions beyond N$^4$LO}
\begin{figure}[t]
  \begin{center}
\includegraphics[width=0.8\textwidth,keepaspectratio,angle=0,clip]{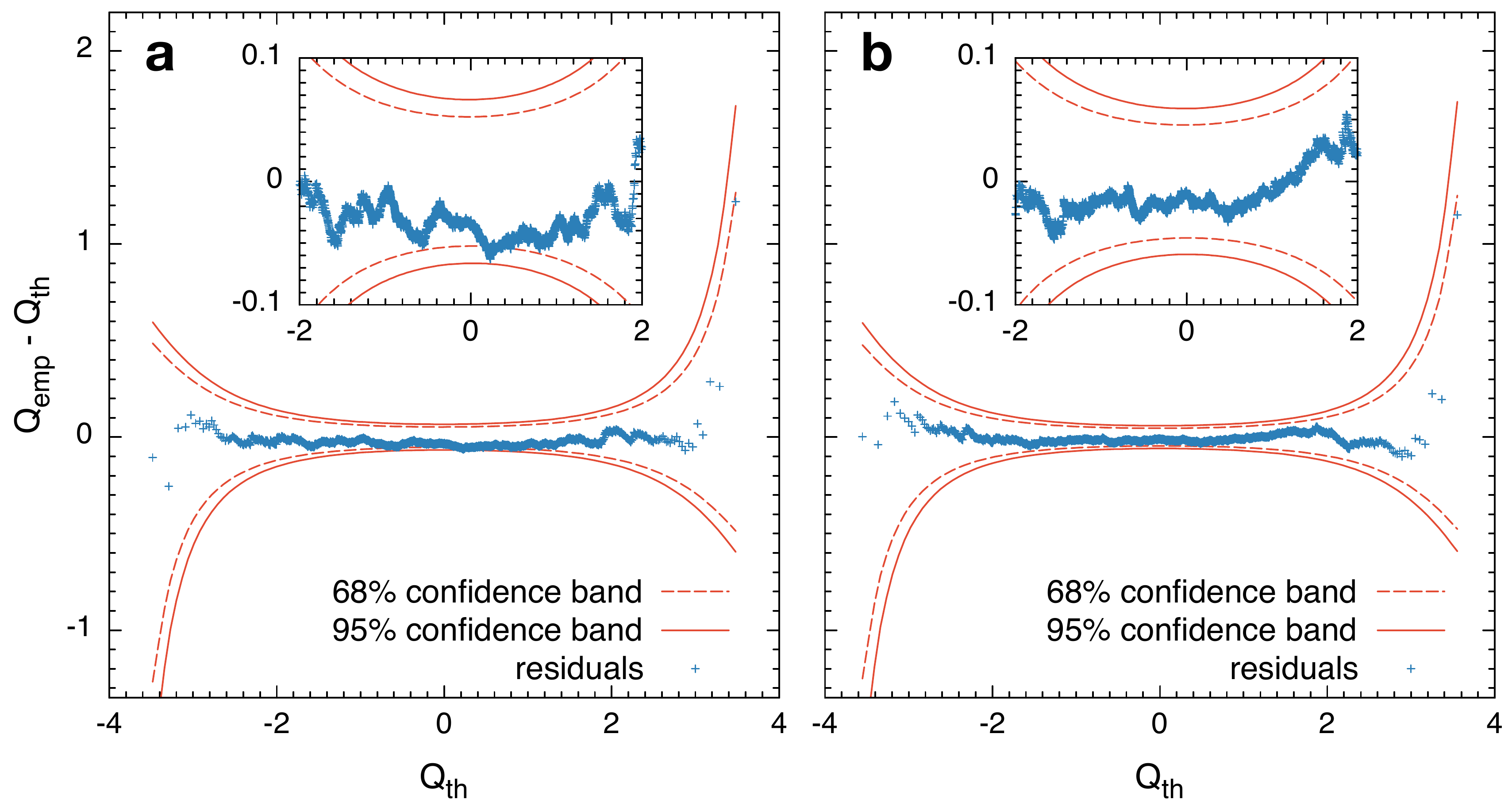}  
\end{center}
\vskip -0.4 true cm
\caption{Tail-sensitive rotated
    quantile-quantile plots for Model 1 and $\Lambda =
      463.5$~MeV.
The plots \textbf{a} and \textbf{b} show the results of the fits in
the energy range of $E_{\rm lab} = 0-220$~MeV and  $E_{\rm lab} =
0-300$~MeV, respectively.  
  The distribution of
  residuals, shown by blue crosses, is consistent with the normal distribution $\mathcal{N}(0,1)$ at a given
  confidence level if all points lie inside the corresponding
  confidence band. Dashed and solid lines mark the $68\%$ and $95\%$
  confidence bands, respectively. 
\label{TStest}
}
\end{figure}

To extract the constants $f^2$ from NN data, one needs to specify the
maximum energy of data considered in the PWA $E_{\rm lab}^{\rm max}$. To stay unbiased, we
perform determinations of $f^2$'s at different values of $E_{\rm
  lab}^{\rm max}$ varied in steps of $20$~MeV. Since our theoretical
framework and the employed parametrization of the scattering amplitude
are valid below the pion
production threshold, the largest value of $E_{\rm
  lab}^{\rm max}$ we consider is $E_{\rm
  lab}^{\rm max} = 300$~MeV. On the other hand, when lowering 
$E_{\rm lab}^{\rm max}$, the number of experimental data
gets reduced leading to larger statistical errors, and one runs the
danger of overfitting. To verify consistency of our fit results at
individual energies, we compare the resulting quantiles of the
empirical distribution of residuals against the ones of the assumed
normal distribution $\mathcal{N}(0,1)$. In order to statistically
quantify deviations from the Gaussian distribution of residuals, various
confidence bands have been derived in the literature. In this work 
we employ the ones by Aldor-Noiman et al.~\cite{Aldor-Noiman:2013} 
which are one of the most recent and most sensitive, especially towards the 
tails of the quantile-quantile plot.   
In Fig.~\ref{TStest}\textbf{b}, one observes that the corresponding
graphical normality test is fulfilled at the $1\sigma$-level for the
fit up to $E_{\rm lab}^{\rm max} = 300$~MeV. When the energy
range of the fit is reduced to $\sim 200$~MeV, we start observing
systematic distortions from the normal distribution of residuals
pointing towards a possible overfitting issue. We, therefore, do not
take into account fits with $E_{\rm lab}^{\rm max}$ below $E_{\rm
  lab}^{\rm max} = 220$~MeV. For this energy range, the distribution
of empirical residuals is still statistically consistent with the
normal one at the $2\sigma$-level, see Fig.~\ref{TStest}\textbf{a}. 
This leaves us with five energy ranges corresponding to $E_{\rm
  lab}^{\rm max} = 220$, $240$, $260$, $280$ and $300$~MeV.

\begin{table}[t]
\caption{Central values and statistical errors of the $\pi$N
    coupling constants $f_i^2$ for all
    considered interaction models and $E_\mathrm{lab}^\mathrm{max}$ values
    \label{tab2}}
\smallskip
\begin{tabular*}{\textwidth}{@{\extracolsep{\fill}}rlllll}
   \noalign{\smallskip}
  \hline
   \noalign{\smallskip}
%  \noalign{\smallskip}
 & $220$ MeV & $240$ MeV & $260$ MeV & $280$ MeV & $300$ MeV \\
 \noalign{\smallskip}
  \hline
   \noalign{\smallskip}
  Model 1: &&&&& \\[1pt]
  $\fp^2$ & $0.0772(5)$ & $ 0.0772(5)$ &  $0.0769(4)$ &  $0.0768(4)$  & $0.0769(4)$ \\[1pt]
  $f_0^2$ & $0.0782(10)$ & $ 0.0783(9)$ &  $0.0781(8)$ &  $0.0780(8)$
                                                        & $0.0782(8)$
  \\[1pt]
  $\fc^2$ & $0.0770(5)$ & $ 0.0770(5)$ &  $0.0769(4)$ &  $0.0767(4)$  & $0.0766(4)$ \\
 \noalign{\smallskip}
  \hline
   \noalign{\smallskip}
  Model 2: &&&&& \\[1pt]
  $\fp^2$ & $0.0772(5)$ & $ 0.0772(5)$ &  $0.0769(4)$ &  $0.0768(4)$  & $0.0769(4)$ \\[1pt]
  $f_0^2$ & $0.0782(9)$ & $ 0.0782(9)$ &  $0.0780(8)$ &  $0.0779(8)$
                                                        & $0.0782(8)$
  \\[1pt]
  $\fc^2$ & $0.0771(5)$ & $ 0.0771(5)$ &  $0.0770(4)$ &  $0.0767(4)$  & $0.0767(4)$ \\
 \noalign{\smallskip}
  \hline
   \noalign{\smallskip}
  Model 3: &&&&& \\[1pt]
  $\fp^2$ & $0.0772(5)$ & $ 0.0772(5)$ &  $0.0770(4)$ &  $0.0768(4)$  & $0.0770(4)$ \\[1pt]
  $f_0^2$ & $0.0778(9)$ & $ 0.0779(9)$ &  $0.0776(8)$ &  $0.0775(7)$
                                                        & $0.0777(7)$
  \\[1pt]
  $\fc^2$ & $0.0772(5)$ & $ 0.0771(5)$ &  $0.0771(4)$ &  $0.0768(4)$  & $0.0768(4)$ \\
 \noalign{\smallskip}
  \hline
\end{tabular*}
\end{table} 

In Table \ref{tab2}, we show the extracted values of the constants $f^2$
for three different interaction models and five energy ranges.
In each case, we have determined the maximum of the marginal posteriors 
$p(f^2 \vert D)$ using the mutually consistent np and pp data from the independent 
data selection up to the given value of $E_{\rm lab}^{\rm max}$ for
Models 1, 2 and 3 by numerically evaluating the integral in Eq.~(\ref{eq:pdf-final}).
Around $50$ evaluations of the marginal posterior turn out to be sufficient to 
accurately determine the maximum of $p(f^2 \vert D)$.
For the employed grid of 8 $\Lambda$-values, 
we thus had to carry out $\sim 400$ determinations of the set of $30$ LECs $C$ 
for every of the $15$ considered cases. 
% 45-55 determinations

While the central values of the $f^2$ can be conveniently determined in this way, 
numerically obtaining the full marginal posteriors on a sufficiently dense grid 
is computationally costly, and it was only done for our central model with 
$E_{\rm lab}^{\rm max} = 280$~MeV as shown in Fig.~2 of the main text, This
required performing $8000$
fits of the contact LECs $C$. As discussed in the text, the marginal posterior 
can be very well approximated by a multivariate Gaussian distribution, whose
central values are given by the maximum of $p(f^2 \vert D)$ and whose covariance
matrix can be inexpensively calculated from the $\sim 50$ points evaluated during 
the determination of the maximum. The covariance matrices then yield the 
uncertainties quoted in Table~\ref{tab2} and, for the sake of completeness, we also
give the corresponding correlation coefficients in Table~\ref{tab3}.
In order to arrive at our final recommended values for the $\pi$N couplings $f_i^2$, 
we perform a simple averaging of the $15$ multivariate Gaussian distributions 
obtained above
% \begin{equation}
%   \label{eq:averaged-pdf}
%   p_\mathrm{avg}(f^2 \vert \bar{D}) = \frac{1}{15}
%   \sum_{i,j} p(f^2 \vert D_i \, M_j) \, ,
% \end{equation}
\begin{align}
  \label{eq:averaged-pdf}
  p_\mathrm{avg}(f^2 \vert \bar{D}) &= \frac{1}{5}\sum_{i,j} p(f^2 \vert D_i \, M_j) \, p(M_j \vert D_i) \, ,
  \nonumber\\
  &\approx \frac{1}{15} \sum_{i,j} p(f^2 \vert D_i \, M_j) \, ,
\end{align}
where $p(f^2 \vert D_i \, M_j)$ is given by Eq.~(\ref{eq:pdf-final}) 
with $D_i \equiv D(E_\mathrm{lab,i}^\mathrm{max})$ being the selected data for 
$E_\mathrm{lab}^\mathrm{max} = 220,240,260,280,300$~MeV and $\{M_j\}$ is the set of the three
potential models. %which both were notationally dropped above.  
The $p(M_j \vert D_i)$ are the model weights for the different models and different values 
of $E_\mathrm{lab}^\mathrm{max}$. Since the description of the data is approximately equal 
among the three models, we have set the $p(M_j \vert D_i) = 1/3$ to be equal as well in the 
second line of Eq.~\eqref{eq:averaged-pdf}.
Furthermore, $\bar{D}$ denotes the dependence on the data averaged over 
$E_\mathrm{lab}^\mathrm{max}$.
% Therefore, 
% the averaging corresponds to both a marginalization
% over the discrete parameter $E_\mathrm{lab}^\mathrm{max}$ (with uniform prior) and 
% a model averaging over Models 1-3. 
In the simple averaging of Eq.~\eqref{eq:averaged-pdf} all
combinations of Models and $E_\mathrm{lab}^\mathrm{max}$ enter with
the same weight. The straightforward averaging over the data for
different $E_\mathrm{lab}^\mathrm{max}$ reflects that all values of
$E_\mathrm{lab}^\mathrm{max}$ are to be treated equally on statistical
grounds. Since the maximum of $p(f^2 \vert D_i \, M_j)$ gets more
localized in the presence of more data (i.e. for higher values of
$E_\mathrm{lab}^\mathrm{max}$) and simultaneously larger in magnitude
(due to the normalization of $p(f^2 \vert D_i \, M_j)$), there is a
natural weighting towards the posteriors including more data. The
development of a formal Bayesian model
for the data averaging over $E_\mathrm{lab}^\mathrm{max}$ goes beyond
the scope of the present study. Given the rather stable results for
the coupling constants $f_i^2$  with
respect to the variation of $E_\mathrm{lab}^\mathrm{max}$, see Tables
\ref{tab2} and \ref{tab3}, we expect the employed procedure
for averaging over $E_\mathrm{lab}^\mathrm{max}$ to be sufficient for the
purpose of this study.
The final results for the central values and 
errors ($\rm a$) of the $f_i^2$ given in Eq.~(4) of the main text have been obtained 
by approximating $p_\mathrm{avg}(f^2 \vert D)$ with a single multivariate 
Gaussian around its maximum.

\subsection{7.~Verification of Bayesian model weights}
In the averaging of Eq.~\eqref{eq:averaged-pdf} used to obtain our recommended values of the pion-nucleon coupling constants, we have set the model weights $p(M_j \vert D_i)$ to be equal. In order to verify this approximation, we have repeated the model averaging with the exact Bayesian model weights
\begin{equation}
  \label{eq:bayesian-model-weights}
  p(M_j \vert D_i) = \frac{p(D_i \vert M_j) \, p(M_j)}{\sum_k p(D_i \vert M_k) \, p(M_k)} \rightarrow \frac{p(D_i \vert M_j)}{\sum_k p(D_i \vert M_k)} \, ,
\end{equation}
where $p(D_i \vert M_j)$ is the data posterior of the data $D_i$ for a particular model $M_j$ and $p(M_j)$ is the model prior. We choose a uniform model prior $p(M_k) = 1/N_M$ with $N_M = 3$ the number of models, yielding the last expression in Eq.~\eqref{eq:bayesian-model-weights}. The data posterior can be obtained by marginalizing over all parameters of the joint posterior
\begin{equation}
  \label{eq:data-posterior}
  p(D_i \vert M_j) = \int df^2 d\Lambda \, dC \; p(D_i \vert f^2 C \Lambda M_j) \, p(f^2 C \Lambda) \, .
\end{equation}
Eq.~\eqref{eq:data-posterior} can be obtained from Eq.~(2) of the main text where the dependence on the model $M_j$ had been notationally suppressed. In a frequentist analysis the model weights of Eq.~\eqref{eq:bayesian-model-weights} are qualitatively similar to a weighting with the exponential of the $\chi^2$ value of each model $M_j$. The model weights thus become important when the involved models do not describe the data equally well. Employing our Bayesian model weights $p(M_j \vert D_i)$ in the first line of Eq.~\eqref{eq:averaged-pdf} and repeating the averaging, we obtain
\begin{eqnarray}
  \label{FinalResultBayesianWeights}
    \fp^2 &=& 0.0770(5)^{\rm a} \,, \nn
    f_0^2 &=& 0.0780(9)^{\rm a} \,,\nn
    \fc^2 &=& 0.0768(5)^{\rm a} \,,
\end{eqnarray}
in good agreement with our recommended values of Eq.~(4) of the main text.

\begin{table}[t]
  \caption{Correlation coefficients between $f_i^2$'s for all
    considered interaction models and $E_\mathrm{lab}^\mathrm{max}$ values
    \label{tab3}}
\smallskip
\begin{tabular*}{\textwidth}{@{\extracolsep{\fill}}rrrrrr}
 \noalign{\smallskip}
  \hline
   \noalign{\smallskip}
%  \noalign{\smallskip}
 & $220$ MeV & $240$ MeV & $260$ MeV & $280$ MeV & $300$ MeV \\
 \noalign{\smallskip}
  \hline
   \noalign{\smallskip}
  Model 1: &&&&& \\[1pt]
  corr($\fp^2$,$f_0^2$) & $-0.05$ & $-0.04$ & $-0.01$ & $-0.04$ & $-0.04$ \\[1pt]
  corr($f_0^2$,$\fc^2$) & $-0.13$ & $-0.08$ & $-0.12$ & $-0.14$ & $-0.15$ \\[1pt]
  corr($\fc^2$,$\fp^2$) &  $0.32$ &  $0.28$ &  $0.35$ &  $0.28$ &  $0.27$ \\
 \noalign{\smallskip}
  \hline
   \noalign{\smallskip}
  Model 2: &&&&& \\[1pt]
  corr($\fp^2$,$f_0^2$) & $-0.04$ & $-0.04$ & $-0.01$ & $-0.04$  & $-0.04$ \\[1pt]
  corr($f_0^2$,$\fc^2$) & $-0.10$ & $-0.09$ & $-0.12$ & $-0.14$  & $-0.12$ \\[1pt]
  corr($\fc^2$,$\fp^2$) &  $0.29$ &  $0.29$ &  $0.35$ &  $0.28$  &  $0.27$ \\
 \noalign{\smallskip}
  \hline
   \noalign{\smallskip}
  Model 3: &&&&& \\[1pt]
  corr($\fp^2$,$f_0^2$) &  $0.00$ &  $0.00$ &  $0.04$ &  $0.06$  &  $0.06$ \\[1pt]
  corr($f_0^2$,$\fc^2$) & $-0.03$ & $-0.01$ & $-0.05$ & $-0.07$  & $-0.04$ \\[1pt]
  corr($\fc^2$,$\fp^2$) &  $0.29$ &  $0.28$ &  $0.33$ &  $0.28$  &  $0.25$ \\
 \noalign{\smallskip}
  \hline
\end{tabular*}
\end{table}

\subsection{8.~Sensitivity to the values of the higher-order $\mathbf{\pi}$N LECs}
To propagate the statistical errors in the values of the higher-order $\pi$N
LECs $c_i$, $\bar d_i$ and $\bar e_i$ determined from the Roy-Steiner equation analysis of
Ref.~\cite{Hoferichter:2015tha_suppl}, see Table \ref{tab1}, we have
generated a sample of $50$ sets of normally distributed values of
these LECs. For each set, the determination of $f^2$ is repeated for the 
central model (albeit with different values of the 
corresponding $\pi$N LECs each time) and $E_\mathrm{max}=280$~MeV. 
The resulting uncertainties of the $f^2$ is then obtained by taking the
standard deviation of the these $50$ values of the $f^2$.
%While the number 
%of sets is restricted by computational cost and rather low for 
%such a Monte Carlo-like sampling, it should give a reasonable estimation
%of the size of the error induced by the statistical uncertainties of the 
%Roy-Steiner analysis.

\subsection{9.~Impact of scattering data after 1993}
In Fig.~3 of the main text we compare our results for the $\pi$N
coupling constants to various other determinations of the last 30
years. Here, the results of the Nijmegen group
\cite{Stoks:1992ja_suppl,Klomp:1991vz_suppl} and the Granada group
\cite{Perez:2016aol_suppl} are of particular interest since they are also
obtained from NN scattering data. As already pointed out in the main
text, we find, similarly to the Granada 2017
analysis, significantly larger values of all coupling
constants $\fp^2$, $f_0^2$ and $\fc^2$ compared to the Nijmegen
results. The database of scattering data used by the Granada 2017
analysis is much more similar to ours than the older Nijmegen analyses
of Refs.~\cite{Stoks:1992ja_suppl,Klomp:1991vz_suppl} which evidently do not
include many experimental data obtained after 1993. This raises the
question if the larger values of the coupling constants are driven by
the differences in databases. 

In order to address this issue, we have repeated our analysis using
scattering data up to 1993 for the Model 1 and
$E_\mathrm{lab}^\mathrm{max} = 280$~MeV.\footnote{We do keep,
  however, the np coherent scattering length of
  Ref.~\cite{Schoen:2003my}. We have checked that this does not affect
  the conclusions of this section.} To avoid possible overfitting
issues due to a significantly smaller number of data and to facilitate
a comparison with Ref.~\cite{Perez:2016aol_suppl}, we restrict ourselves 
to the average value of the coupling constant by setting 
$f_p^2 = f_c^2 = f_0^2 \equiv f^2$ in the NN interaction.
Performing a fit of the np
and pp data using the complete 2020 database with $N_{\rm dat} = 4926$, we
obtain the average value of the pion nucleon coupling constant of
$f^2 = 0.07694(29)$ in a good agreement with the corresponding
values quoted in Table \ref{tab2}. 

We now repeat the determination by retaining only the experimental data available in
1993.  The data selection using the
$3\sigma$-criterion is mostly consistent with our final 2020 database
restricted to data before and including 1993. Concerning the
differences, we find that the proton-proton differential cross
sections of Refs.~\cite{Cox:1968jxz,Jarvis:1971fla,PALMIERI1958299}
rejected in Table~\ref{tab:pp-data-differences} are now included while
the previously included dataset at $E_\mathrm{lab} = 137$~MeV of
Ref.~\cite{PALMIERI1958299} is instead rejected. In the
neutron-proton database, the dataset at $E_\mathrm{lab} = 96$~MeV of
Ref.~\cite{Griffith_1958} is additionally rejected due to a too low
$\chi^2$. Using this 1993 database with $N_{\rm dat} = 3804$,
we obtain essentially the same value of the coupling constant,
$f^2 = 0.07698(35)$. Thus, according to our analysis, the
significantly smaller value of $f^2$ found by the Nijmegen group cannot
be explained by the scattering data published after 1993. 
For the sake of completeness, we mention that the Granada
group found slightly smaller values of $f^2$ by excluding the
data published after 1990 and 1995 from the Granada 2013 database\footnote{Notice that both
  the Nijmegen and Granada groups also include scattering data above
  the pion production threshold up to $E_\mathrm{lab} = 350$~MeV in
  their fits.} (without repeating the
data selection), see Table X of Ref.~\cite{Perez:2016aol_suppl}, with the observed
reduction in $f^2$ being, however, by far insufficient to explain the
differences to the much smaller Nijmegen values. 

%Repeating our determination of the $\pi$N coupling constants we find
%$\fp^2 = 0.0767(5)$ and $\fc^2 = 0.0766(5)$ in good agreement with our
%2020 results for Model 1 and $E_\mathrm{lab}^\mathrm{max} = 280$~MeV
%in Table~\ref{tab2}. For $f_0^2$ we obtain a somewhat larger value of
%$f_0^2 = 0.0788(10)$. This shows that the larger values for the
%coupling constants in our analysis (compared to the Nijmegen analysis)
%are not a result of the scattering data published after 1993. 

%\putbib[supplemental_sorted]
%\end{bibunit}
%Control: production of eprint (1) enabled
%

\end{document}